\documentclass{icrc}
\usepackage{times}
\usepackage{graphicx}
\def\la{\hbox{{\lower -2.5pt\hbox{$<$}}\hskip -8pt\raise
-2.5pt\hbox{$\sim$}}}
\def\ga{\hbox{{\lower -2.5pt\hbox{$>$}}\hskip -8pt\raise
-2.5pt\hbox{$\sim$}}}

\begin{document}

\title{WIMPs Are Stronger When They Stick Together}
\author[1,3]{A. V. OLINTO}
\affil[1]{Department of Astronomy \& Astrophysics,  \& Enrico Fermi
Institute, The University of Chicago, Chicago, IL 60637}
\author[2,3]{P. BLASI}
\affil[2]{Fermi National Accelerator Laboratory, Batavia, IL
60510-0500}
\affil[3]{Arcetri Osservatorio, Firenze, Italia}
\author[1]{C. TYLER}

\correspondence{olinto@oddjob.uchicago.edu}

\firstpage{1}

\pubyear{2001}


\maketitle

\begin{abstract}
Weakly interacting massive particles (WIMPs) remain the
strongest candidates for the dark matter in the Universe. If WIMPs are
the dark matter, they will form galactic halos
according to the hierarchical clustering observed in N-body simulations.
Cold dark matter (CDM) simulations  show that
large dark matter structures such as galactic and cluster halos are
formed by the merging of many smaller clumps of dark matter.  Each
clump or halo is characterized by a centrally cusped density profile
that can enhance the rate of WIMP annihilation and make the annihilation
products more easily detectable. Electrons and positrons generated as
decay products of WIMP annihilation emit synchrotron radiation in the
Galactic magnetic field.  We study the synchrotron signature from the
clumps of dark matter in our Galactic halo. We find that the emission in
the radio and  microwave region of the electromagnetic spectrum
can be above the CMB anisotropy level and should be detectable by CMB
anisotropy experiments. Depending on the density profile of
dark matter clumps, hundreds of clumps can have detectable fluxes and
angular sizes.
\end{abstract}

\section{Introduction}

The density of dark matter in the present Universe is observed via its
gravitational effects on galaxies and clusters of galaxies to
constitute at least about 30\% of the critical density of the
Universe, but the  nature of this dark matter is still unknown. Primordial
nucleosynthesis constrains the density of baryonic matter to be less than
about 5\% of the critical density, thus most of the dark matter is
non-baryonic. The leading candidate for the dark matter is the lightest
supersymmetric particle in supersymmetric extensions
of the standard model that is stable by conservation of R-parity.
In most scenarios this weakly interacting massive particle (WIMP)
is the neutralino, $\chi$ (for a review, see \citet{jkg96}).

Neutralinos may be detected directly as they traverse the Earth or
indirectly by the observation of their annihilation products. Direct
neutralino searches  are now underway in a number of low
temperature experiments with no consensus detection as of yet.
Indirect searches have been proposed both for gamma rays  and
synchrotron emission from the annihilation of WIMPs in the Galactic center
\citep{bgz92, bbm94} where the WIMP density and magnetic field around a
central massive black hole may enhance the emission
significantly \citep{gs99,g00}. The rate of annihilation is
proportional to the neutralino density squared ($\sim n_{\chi}^2$),
therefore the strongest flux is expected to come from the Galactic
center where  the dark matter halo density  peaks.

  N-body cold dark matter (CDM) simulations have shown
that the dark matter halos have a  density
profile with a cusp at the center  (within a core radius $R_c \sim 5 -
10$ kpc) and a steeper profile in the outer regions. The slope of the
inner cusp is still a matter of debate ranging from $r^{-1}$ to $r^{-2}$
in different simulations.
Superimposed on the smooth component, the high-resolution
simulations find a large degree of substructure formed due to the
constant merging of smaller halos
to form the present dark matter halo  (see, for example,
\citet{gmglqs}). The large number of clumps generated through the
hierarchical clustering of dark matter comprise about $\sim 10-20\%$ of
the total mass and can enhance significantly the emission of
gamma rays and neutrinos from neutralino annihilation in higher
density clumps \citep{begu99,cm01}.  We show here (and in more detail in
\citet{bot01}) that the synchrotron radiation of electrons and positrons
generated as decay products of WIMP annihilation in the Galactic magnetic
field can provide a crucial test of WIMP models, since the predicted
fluxes are in the microwave region and exceed the signal of CMB
anisotropies at some frequencies. The detection of this excess radiation
from small angular size regions in the sky may provide the first signal
from WIMP annihilations.

\section{Synchrotron Signature of WIMP Clumps}

The annihilation of neutralinos produces high energy particles through
several processes, depending on the mixture of supersymmetric fields that
form the neutralino. In addition to the gamma ray line generated in the
channel $\chi {\bar \chi} \to \gamma \gamma$, the annihilation of two
neutralinos also results in a continuum of particles (gamma rays,
neutrinos, electrons, positrons, muons, etc.) that have energy spectra
well represented by an $E^{-3/2}$ power law. This is typical of the
process of fragmentation and hadronization of quarks into hadrons (mainly
pions) and their decay into secondary products. Here we concentrate on
the process $\chi {\bar \chi} \to q {\bar q}  \to$ hadrons and the
decay of the resulting pions. (We neglect the small contribution
from kaons and other mesons.) In particular, we are interested
in the $e^+e^-$ pairs generated by the decay of the charged pions.

When the annihilation of neutralinos occurs in the Galaxy,
the secondary products are injected in the ambient magnetic
field.  Clumps can be  brightened by the synchrotron emission of
$e^+e^-$ pairs in the Galactic field, depending on their
position in the halo.

We initially assume that the electrons emit in the same region
in which they are generated, which is the case when $e^{\pm}$ are
magnetically constrained in the higher magnetic field regions. As the
field decays, the diffusion of the emitting $e^{\pm}$ needs to be
included (see \citet{bot01}).

The spectrum of the generated $e^{\pm}$ naturally cuts off at about the
neutralino mass, $ m_{\chi}$. Therefore, the relevant frequency range for
$e^{\pm}$ synchrotron emission lies below the maximum frequency,
\begin{equation}
\nu_{\rm max} \simeq \ B_{\mu} (m_{\chi}/ {\rm 100 \ GeV})^2 \, {\rm GHz} \ ,
  \end{equation}
where
$B_{\mu} = B/ \mu$G. Since the Galactic magnetic field is around
$\mu$G, for $m_{\chi}\  \ga \ 100$ GeV, the radiation
will extend up to microwave frequencies.  Note that the
electron-positron spectra are flatter than $E^{-2}$ which implies that
most of the energy is carried by the most energetic $e^{\pm}$ particles,
and most of the synchrotron emission occurs at frequencies approaching
the cutoff.

\begin{figure}[t]
  \vspace*{2.0mm}
\includegraphics[angle=-90,width=8.3cm]{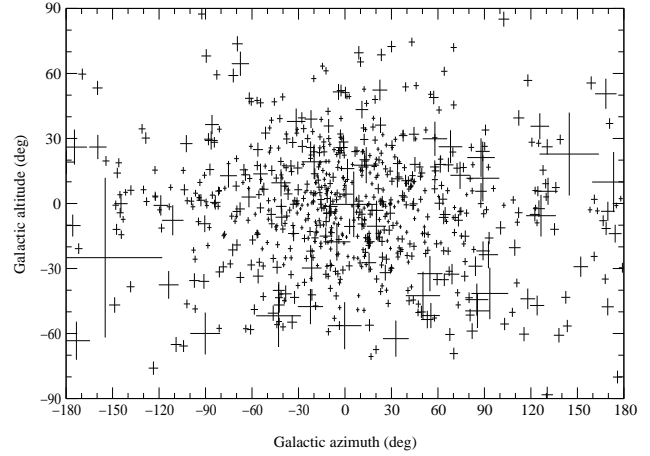}
  \caption{Locations for 754 dark matter clumps with
fluxes higher than CMB anisotropies between 1 and
1000 GHz. This realization of the Galactic halo follows a NFW profile
and  contains 3972 total clumps with $M_{cl} \ge 10^7 M_{\odot}$.
  Crosses represent the solid angle of each
clump (multiplied by a factor
of 5)  inside which 90\% of the flux is located.}
  \end{figure}

In our calculations, we assumed that the
Galactic magnetic field has an exponential scale height  as in
\citet{st97}.  To model the smooth halo component, we considered
the NFW halo profile
\citep{nfw}
\begin{equation}
\rho_{\rm halo} = m_{\chi} n_0 \left({r \over r_c}\right)^{-1} \left[{1 +
{r\over r_c}}\right]^{-2}  \ ,
\end{equation}
where $r_c$ is the core radius and $n_0$ is the number density at
$r_c$.  The two parameters, $r_c$ and $n_0$, can be set by requiring that
the halo contains a given total mass ($M_H$) and that the velocity
dispersion at some distance from the center is known (in the case of the
Galaxy,  the velocity dispersion is $\sim 200$ km/s in the vicinity of
our  solar system).

We modeled the clumpy halo following
\citet{bs00} where they fit the simulations to a joint distribution of
clump mass, $m$, and position, $r$, by
\begin{equation}
n_{cl}(r,m) = n_{cl,0} \left(\frac{m}{M_H}\right)^{-\alpha}
\left[1+\left(\frac{r}{r_c^{cl}}\right)^2\right]^{-3/2},
\end{equation}
where $n_{cl,0}$ is a normalization constant, $r_c^{cl}$ is the
core of the clumps distribution, and $\alpha\sim 1.9$ fits
well the simulations in \cite{gmglqs}.
  In \cite{gmglqs}, a halo with $M_H\approx 2\times 10^{12}~M_\odot$
contains about $500$ clumps with mass larger than $M_{cl} \sim
10^8~M_\odot$.
We present here the results for the case of dark matter halos following a
NFW profile while other cases are considered in detail in  \citet{bot01}.
The density of each clump is taken to be as in Eq. (2), where the
normalization constant is calculated from the total mass, the core radius
of the clump is assumed to be 0.1 of the clump radius and the latter is
taken from the condition that the clump density equals the local density
of dark matter in the Galaxy at the clump position.

We simulated several realizations of a clumpy halo each with about 4000
clumps of masses $M_{cl} \ge 10^7 M_{\odot}$.
Figure 1 shows one realization with 3972 clumps in Galactic coordinates.
Here we adopt a cross section for neutralino annihilation
$\langle \sigma v \rangle_{\chi {\bar \chi}} = 3 \times 10^{-29} \,
{\rm cm^3/s}$.
The results can be rescaled for the present choice of
dark matter density profile (but this rescaling is not generic, see
\citet{bot01}).

In Figure 1, 754 clumps have synchrotron fluxes in the range 1 to 1000
GHz above $10^{-5}$ times the CMB flux. These clumps are potentially
detectable by anisotropy experiments where the isotropic CMB emission is
subtracted and only anisotropies at the level of  $10^{-5}$ times the CMB
flux remain.   Of these, 132 are above 30 degrees Galactic latitude (or
below -30).  The clumps range in angular sizes from the size of a pixel
(10'x10') to about 10 degrees, occupying 2\% of the solid angle of the sky.
The crosses shown in the figure represent the solid angle (amplified by a
factor of 5 for clarity) inside which 90\% of the radiation for each
clump is located. Figure 2  shows the location and solid angles of clumps
that are observable between 10  and 400 GHz. This range
is expected to be relatively quiet of CMB foregrounds (see, e.g.,
\citet{teho00}), therefore the most sensitive CMB anisotropy experiments
are planned for this range in frequency.

  \begin{figure}[t]
  \vspace*{2.0mm}
  \includegraphics[angle=-90,width=8.3cm]{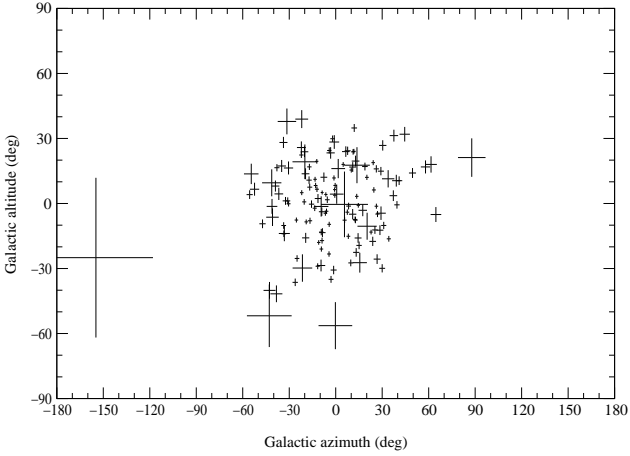}
  \caption{Same as Fig. 1 but for observable clumps with fluxes higher
than CMB anisotropies between 10 and 400 GHz.}
  \end{figure}

The histogram in Fig. 3 shows  the number of  observable clumps  of different
solid angles.
There are over 100 clumps that can be observed by experiments with angular
resolution around 0.2 degrees.  A few very large objects can be seen even
by experiments with poor angular resolution. These correspond to clumps
located close to Earth.  If a large object is identified in this
frequency range, a spectral study would verify its nature as a dark
matter clump. In addition, the radial dependence of the flux within the
clump could further constrain
  CDM clustering behavior.

Figure 4 shows a histogram of the number of observable clumps with total
flux in or above a given energy bin. The histograms in Figs. 3 and 4 were
generated for the same realization in Fig. 1. For different realizations
see \citet{bot01}. The behavior shown in these figures is generic and
can act as a guide to determine if microwave sources
are annihilating dark matter clumps or some other foreground.

  \begin{figure}[t]
  \vspace*{2.0mm}
  \includegraphics[angle=-90,width=8.3cm]{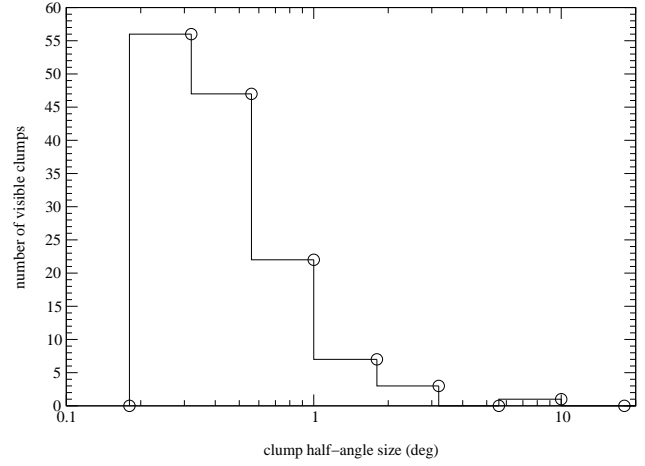}
  \caption{Number of clumps of each angular size with fluxes above CMB
anisotropies  between 10  and 400 GHz. }
  \end{figure}

  \begin{figure}[t]
  \vspace*{2.0mm}
  \includegraphics[angle=-90,width=8.3cm]{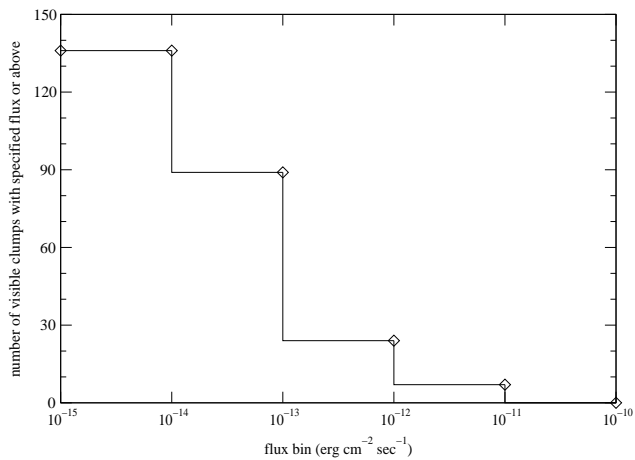}
  \caption{Number of observable clumps with total flux in the specified
flux bin or above.}
  \end{figure}

As an example of the spectral dependence of the synchrotron emission
from WIMP annihilation, we show in  Figure 5  the spectrum of a dark
matter clump (thick line) compared with the CMB anisotropies
(thin line) in the same solid angle occupied by the clump. This clump is
chosen to lie at galactocentric coordinates [-4,0,0] kpc (where the Sun is
located at [-8.5,0,0]), with $10^8 M_{\odot}$, occupying a half-angle of
1 degree on the sky.  The neutralino mass was chosen to be 100 GeV.
The dashed line  shows the case of a neutralino with 10 TeV for
comparison. The cutoff is moved to higher frequencies as expected from
Eq. (1). If an annihilating dark matter clump were to be observed,
the cutoff would give us the neutralino mass directly.  The dotted
line shows the flux for the same clump  with $m_{\chi} = 100$
GeV but with the  interaction
strength, given by the WIMP annihilation cross section times
velocity,  $\langle \sigma v \rangle_{\chi {\bar \chi}}$,
increased by a factor of 100.

  \begin{figure}[t]
  \vspace*{2.0mm}
  \includegraphics[angle=-90,width=8.3cm]{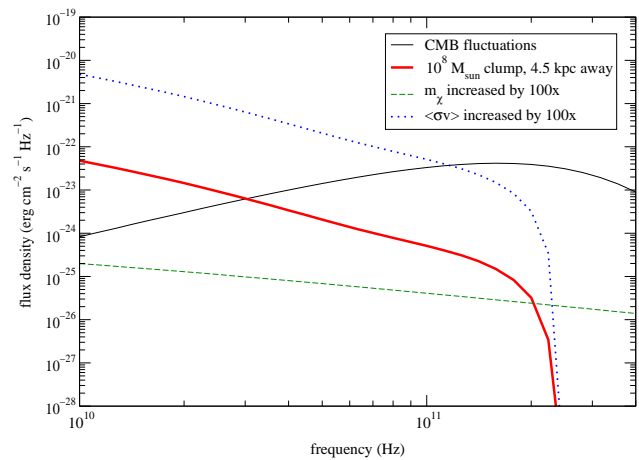}
  \caption{Spectrum of synchrotron emission from an example NFW
dark matter clump with $m_{\chi} = 100$ GeV  (thick line) compared with
the CMB anisotropies (thin line) in the same solid angle occupied by the
clump. This clump is chosen to lie at galactocentric
coordinates [-4,0,0] kpc, with $10^8 M_{\odot}$, occupying a half-angle of
1 degree on the sky.  The flux from the same clump but with $m_{\chi} =
10$ TeV is represented by the dashed line and the dotted line is for
$m_{\chi} = 100$ GeV and
$\langle \sigma v \rangle_{\chi {\bar \chi}}$ increased by
a factor of 100.  }
  \end{figure}

\section{Conclusions}

The clumpy nature of  CDM halos can be used to detect and constrain WIMP
candidates for the non-baryonic dark matter. Neutralino annihilation in
the higher density clumps can be observed via the
synchrotron radiation of electrons and positrons as these annihilation
products radiate in the Galactic magnetic field. The spatial structure
of the Galactic magnetic field  implies that the synchrotron emission
from annihilation gets stronger as clumps get closer to the Galactic
plane. This behavior gives a different angular distribution than the
distribution  from the gamma ray signature of the same clumps. This
unique combination will  help distinguish WIMP clumps from
other extragalactic gamma ray sources.  The frequency range where
these clumps are better observed overlaps with highly sensitive
experiments planned for CMB anisotropy measurements.  The
possibility of detecting these clumps will be soon within reach.
Depending on the density profile of dark matter clumps, hundreds of
clumps have detectable fluxes and angular sizes. Even more
clumps may be present if the lower limit on clump masses is lowered from
$ 10^7 M_{\odot}$.

The spectral shape, spatial distribution, and angular
size of annihilating neutralino clumps discussed above represent some
particular choices of $m_{\chi}$ and
$\langle \sigma v \rangle_{\chi {\bar \chi}}$
which are hard to constrain a priori. The neutralino mass sets
the cutoff of the spectrum and changes the overall flux while the cross
section mostly influences the flux amplitude.   Finally, the
Galactic magnetic field structure above and below the plane of the Galaxy
is poorly known and will also influence the exact observable clump
distribution.  The best strategy is to search for varying sizes of
CMB foregrounds at a number of frequencies and select for those with the
spectral dependence given in Fig. 5. Once some extended synchrotron
sources have been selected, the particular radial distribution and flux
shape will help determine if these sources are dark matter clumps. The
combination of these synchrotron measurements with the direct gamma ray
and neutrino signature will make these sources unique. In addition, the
synchrotron signature will help determine the structure of the magnetic
field above the Galactic disk.

Future CMB experiments such as  MAP and Planck can be used in conjunction
with future gamma ray and neutrino experiments. Full sky coverage helps
this determination since the Galactic plane is usually avoided by small
area experiments. MAP will observe above about 20 GHz while Planck should
start at 30 GHz, with a large increase in angular resolution that
makes these objects easier to detect.

\begin{acknowledgements}

This work was supported by the NSF through grant
AST-0071235 and DOE grant DE-FG0291  ER40606 at the University of
Chicago, and at Fermilab by DOE and NASA grant NAG 5-7092.

\end{acknowledgements}

\end{document}